\documentclass[12pt]{article}

\usepackage{amsmath}
\usepackage{amsthm}
\usepackage{amssymb}
\usepackage{authblk}
\usepackage{easytable}
\usepackage{fullpage}
\usepackage{graphicx}
\usepackage{natbib}
\usepackage[usenames,svgnames]{xcolor}
\usepackage{subcaption}
\usepackage{tikz}
\usepackage[pdftex,
            colorlinks=true,
            linkcolor=NavyBlue,
            citecolor=NavyBlue,
            urlcolor=NavyBlue,
            breaklinks=true,
            backref=page]{hyperref}

\usetikzlibrary{arrows.meta}
\usetikzlibrary{matrix}

\newtheorem{theorem}{Theorem}
\newtheorem{lemma}[theorem]{Lemma}
\newtheorem{definition}[theorem]{Definition}

\newcommand{\N}{\mathbb{N}}  
\newcommand{\Z}{\mathbb{Z}}  

\DeclareMathOperator{\PCTR}{PCTR}

\title{Auction for Double-Wide Ads}

\author[1]{Jonathan Gu}
\author[2]{D\'avid P\'al}
\author[1]{Kevin Ryan}
\affil[1]{Instacart, San Francisco, USA}
\affil[2]{Instacart, New York, USA}
\affil[1,2]{\href{mailto:jonathan.gu@instacart.com,david.pal@instacart.com,kevin.ryan@instacart.com}{\texttt{\{jonathan.gu,david.pal,kevin.ryan\}@instacart.com}}}

\begin{document}

\maketitle

\begin{abstract}
We propose an auction for online advertising where each ad occupies either one
square or two horizontally-adjacent squares of a grid of squares. Our primary
application are ads for products shown on retail websites such as Instacart or
Amazon where the products are naturally organized into a grid. We propose
efficient algorithms for computing the optimal layout of the ads and pricing of
the ads. The auction is a generalization of the generalized second-price (GSP)
auction used by internet search engines (e.g. Google, Microsoft Bing, Yahoo!).
\end{abstract}

\section{Introduction}
\label{section:introduction}

\begin{figure}
\begin{center}
\includegraphics[height=0.9\textheight]{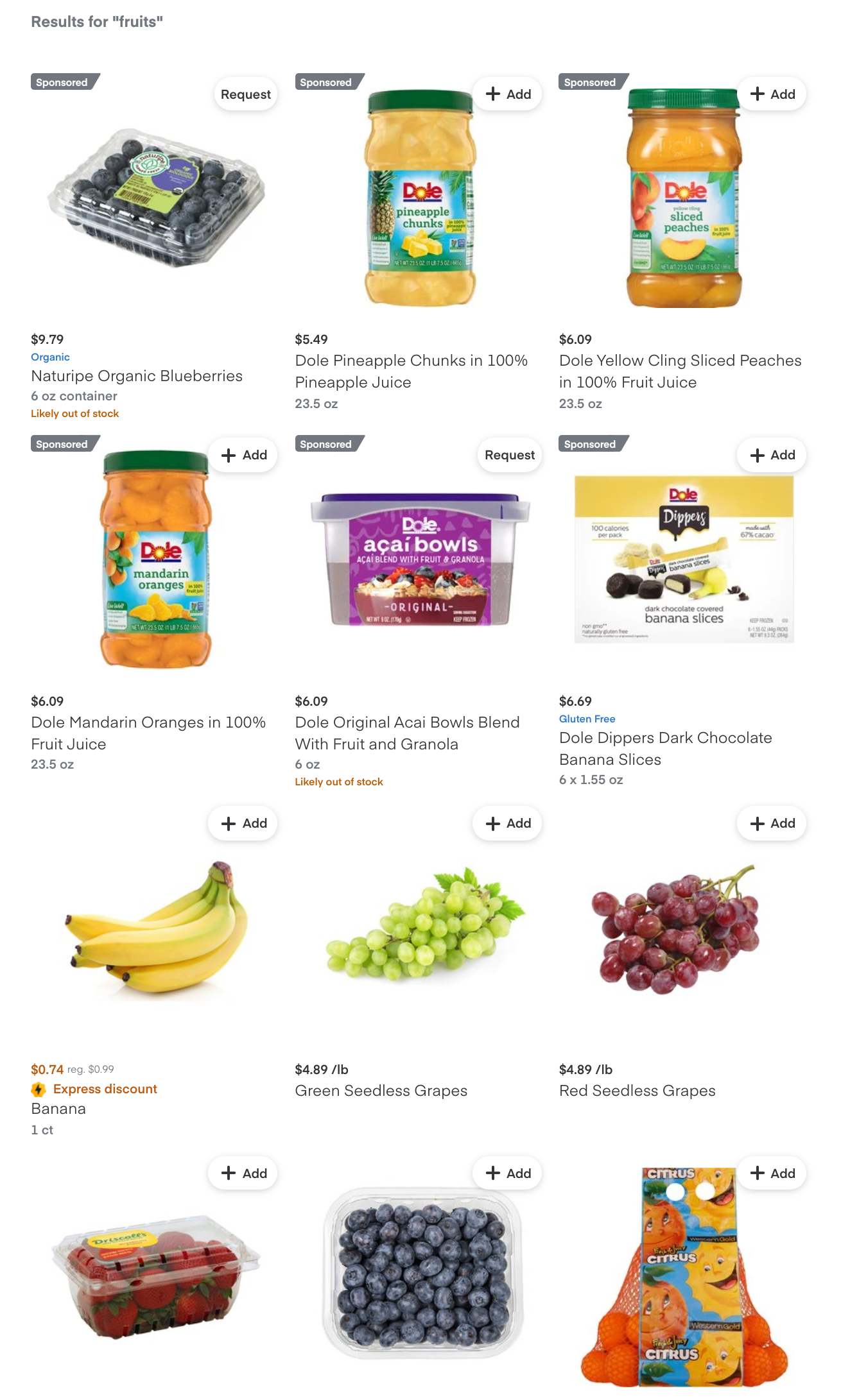}
\caption{\label{figure:products-grid} Top $12$ products shown at Instacart
website for search query ``fruits''. The products are organized in a square grid
with $3$ columns and $4$ rows. The number of the columns in the grid depends on
the width of the screen of the user's device. The sponsored products have a grey
``sponsored`` label attached to them.}
\end{center}
\end{figure}

Online shopping sites, such as Amazon and Intacart, show on their website the
products for sale in a square grid; see Figure~\ref{figure:products-grid}.  The
number of rows and columns of the grid depend on the screen size of user's
device. The products shown and their ordering depends on various contextual
features, e.g., user's search query, department name, and past user's behavior
on the website. Some of the displayed products are sponsored and they are
referred to as \emph{ads}. For legal, regulatory and business reasons, the ads
are distinguished from non-sponsored products by a ``sponsored'' label. For the
same reasons, the ads are retrieved and ordered by a system that is separate
from the system for retrieving and ordering the non-sponsored products.

The selection and the layout of the ads on web page is determined by an auction.
The auction receives as an input the bids from advertisers. The bid expresses
the maximum price the advertiser is willing to pay per click\footnote{Other
payment schemes, such as pay per impression and pay per conversion can be easily
accommodated as well. For ease of exposition, in this paper, we focus solely on
the pay-per click payment scheme.} on their ad. The auction also receives the
predictions of the probabilities that the user clicks on the ads
(\emph{predicted click-through rates}). Based on these inputs, the auction
selects the ads to be shown, determines their layout on the web page, and
computes the prices the advertisers pay per click. If the user clicks on the on
ad, the advertiser is charged the per-click price that was computed by the
auction. This idea goes back to the sponsored search
auctions~\citep{Varian-2007, Varian-2009, Aggarwal-Goel-Motwani-2006,
Edelman-Ostrovsky-Schwarz-2007, Aggarwal-2005,
Aggarwal-Feldman-Muthukrishnan-2006}.

Compared to the sponsored search auctions, grid design offers unique
opportunities. While traditionally each ad occupies one square of the grid, in
this paper, we generalize the design and we consider layouts, in which some of
the sponsored products can occupy two horizontally-adjacent squares of the grid.
Such sponsored products are called \emph{double-wide}. The standard ads
occupying one square are called \emph{single-slot}. The reason for the
double-wide ads is that they can offer more attractive imagery and graphics,
attracting more users to click on them.

In order to accommodate the more general layout, the inputs to the auction
algorithms need to be augmented. The inputs include the the dimensions of the
grid, the list of already occupied squares (occupied by non-sponsored products
or some other piece of content), widths of the ads, and position multipliers.
The position multipliers are similar to the ones used in the truthful variants
of the search ad auctions~\citep{Aggarwal-Goel-Motwani-2006} that arise from the
application of the Vickrey-Clarke-Groves (VCG) mechanism~\citep[Chapter
9.3.3]{Nisan-Roughgarden-Tardos-Vazirani-2007}. However, instead of a single
sequence of position multipliers, we use two sequences, one sequence for
single-slot ads and one sequence for double-wide ads.

We design efficient algorithms for computing the optimal layout of the ads and
their per-click prices. The algorithm for computing optimal layout is based on
dynamic programming. Our algorithm for computing prices is a generalization of
the generalized second price auction (GSP)~\cite[Chapter
28]{Nisan-Roughgarden-Tardos-Vazirani-2007}. We call these GSP-like
prices. As an alternative, one can compute the VCG prices
by a straightforward application of the VCG mechanism. We also design two
extensions of the basic algorithms. The first extension is for the setting where
the auction automatically chooses between a single-slot ad and a double-wide ad
for the same product. The second extension is for a setting where multiple ads
from the same advertiser can be shown on the same web page.

Our work is related to other combinatorial ad auctions. Perhaps the closest is
the work of \cite{Cavallo-Krishnamurthy-Sviridenko-Wilkens-2017}. They
considered the problem of selecting different truncations of text ads for search
advertising on Yahoo!

\subsection{Structure of the paper}

In Section~\ref{section:click-through-rate-model}, we introduce the
click-through rate model that we will use throughout the paper. The model is a
generalization of the separable model introduced by
\cite{Aggarwal-Goel-Motwani-2006}. We also discuss two alternative models. The
alternative models happen to be special cases of our main model.

In Section~\ref{section:layout-algorithm}, we explain the algorithm for
computing the optimal layout. In this section, we also explain  the inputs to
the algorithm. The algorithm is based on dynamic programming. Both the time and
the space complexities are $O(N_1 N_2)$ where $N_1$ is the number of single-slot
ads and $N_2$ is the number of double-wide ads.

In Section~\ref{section:pricing-algorithm}, we explain the algorithm for
computing GSP-like prices. The algorithm builds on the algorithm for computing
the optimal layout. We briefly mention how to compute VCG prices as well.

In Section~\ref{section:extensions}, we explain two extensions of the basic
algorithms. The first extension is for the setting where the auction
automatically chooses between a single-slot ad and a double-wide ad for the same
product. The second extension is for a setting where multiple ads from the same
advertiser can be shown on the same web page.

\section{Click-through rate model}
\label{section:click-through-rate-model}

In this section, we describe the model for the probability that a user clicks on
an ad. This probability is referred to as the \emph{click-through rate}. In
principle, the click-through rate could depend on presence of the other ads and
their positions. However, we assume that the click-through rate of an ad depends
only on the ad itself and the position on the page where the ad is shown. In
particular, we assume that the click-through rate does \emph{not} depend on the
other ads present on the web page.

The position of an ad is specified by its position in the grid. We denote by $R$
and $C$ the number of rows and columns of the grid, respectively. We denote by
$S=RC$ the total number of squares. We number the squares $1,2,\dots,S$ in
row-major order; see Figure~\ref{figure:square-row-major-numbering}. We number
the potential ads that can be shown on the web page by numbers from
$1,2,\dots,N$. Some of these ads are single-slot and some of them are
double-wide. We say that an ad is \emph{placed at position $j$} if either the ad
is single-slot and it covers the square $j$, or the ad is double-wide and it
occupies squares $j$ and $j+1$. We denote by $\PCTR_{i,j}$ the click-through
rate of ad $i$ placed at position $j$.

\begin{figure}
\begin{center}

\begin{tikzpicture}

\node at (1,-1) [minimum size=1cm,draw,ultra thick] (1) {1};
\node at (2,-1) [minimum size=1cm,draw,ultra thick] (2) {2};
\node at (3,-1) [minimum size=1cm,draw,ultra thick] (3) {3};
\node at (4,-1) [minimum size=1cm,draw,ultra thick] (4) {4};
\node at (5,-1) [minimum size=1cm,draw,ultra thick] (5) {5};
\node at (6,-1) [minimum size=1cm,draw,ultra thick] (6) {6};

\node at (1,-2) [minimum size=1cm,draw,ultra thick] (7) {7};
\node at (2,-2) [minimum size=1cm,draw,ultra thick] (8) {8};
\node at (3,-2) [minimum size=1cm,draw,ultra thick] (9) {9};
\node at (4,-2) [minimum size=1cm,draw,ultra thick] (10) {10};
\node at (5,-2) [minimum size=1cm,draw,ultra thick] (11) {11};
\node at (6,-2) [minimum size=1cm,draw,ultra thick] (12) {12};

\node at (1,-3) [minimum size=1cm,draw,ultra thick] (13) {13};
\node at (2,-3) [minimum size=1cm,draw,ultra thick] (14) {14};
\node at (3,-3) [minimum size=1cm,draw,ultra thick] (15) {15};
\node at (4,-3) [minimum size=1cm,draw,ultra thick] (16) {16};
\node at (5,-3) [minimum size=1cm,draw,ultra thick] (17) {17};
\node at (6,-3) [minimum size=1cm,draw,ultra thick] (18) {18};

\node at (1,-4) [minimum size=1cm,draw,ultra thick] (19) {19};
\node at (2,-4) [minimum size=1cm,draw,ultra thick] (20) {20};
\node at (3,-4) [minimum size=1cm,draw,ultra thick] (21) {21};
\node at (4,-4) [minimum size=1cm,draw,ultra thick] (22) {22};
\node at (5,-4) [minimum size=1cm,draw,ultra thick] (23) {23};
\node at (6,-4) [minimum size=1cm,draw,ultra thick] (24) {24};

\end{tikzpicture}

\caption{\label{figure:square-row-major-numbering} The figure shows an example
of a grid with $R=4$ rows and $C=6$ columns. Altogether, the grid has $S=RC=24$
squares. The rows are numbered $1,2,\dots,24$ in row-major order.}
\end{center}
\end{figure}

We generalize the separable click-through rate model
of \cite{Aggarwal-Goel-Motwani-2006} as follows.

\begin{definition}[Separable click-through rates]
\label{definition:separable-click-through-rates}
We assume that there are $3$ sequences of positive real numbers $\alpha_1,
\alpha_2, \dots, \alpha_N$, $\beta_1 \ge \beta_2 \ge \dots \ge \beta_S$, and
$\gamma_1 \ge \gamma_2 \ge \dots \ge \gamma_{S-1}$ such that the click-through
rate of ad $i$ placed at position $j$ is
\begin{equation}
\label{equation:separable-model}
\PCTR_{i,j} =
\begin{cases}
\alpha_i \beta_j & \text{if ad $i$ is single-slot,} \\
\alpha_i \gamma_j & \text{if ad $i$ is double-wide.}
\end{cases}
\end{equation}
The numbers $\alpha_1, \alpha_2, \dots, \alpha_N$ are called \emph{ad factors}.
The numbers $\beta_1, \beta_2, \dots, \beta_S$ and $\gamma_1, \gamma_2, \dots,
\gamma_{S-1}$ are called \emph{position multipliers}.
\end{definition}

Two alternative definitions are possible. The first alternative is to assume
that
\begin{equation}
\label{equation:separable-model-position-based}
\PCTR_{i,j} = \alpha_i \beta_j
\end{equation}
regardless of whether ad $i$ is single-slot or double-wide.
The second alternative is to assume that
\begin{equation}
\label{equation:separable-model-slot-based}
\PCTR_{i,j} =
\begin{cases}
\alpha_i \beta_j & \text{if ad $i$ is single-slot,} \\
\alpha_i (\beta_j + \beta_{j+1}) & \text{if ad $i$ is double-wide.}
\end{cases}
\end{equation}
Clearly, Definition~\ref{definition:separable-click-through-rates} is more
general than either alternative. That is, if click-through rates have the form
\eqref{equation:separable-model-position-based} or
\eqref{equation:separable-model-slot-based}, there exist some (other) sequences
$\alpha_1, \alpha_2, \dots, \alpha_N$, $\beta_1 \ge \beta_2 \ge \dots \ge
\beta_S$ and $\gamma_1 \ge \gamma_2 \ge \dots \ge \gamma_{S-1}$ such that
$\PCTR_{i,j}$ satisfies \eqref{equation:separable-model}. One could argue that
either of the alternatives is more natural than
\eqref{equation:separable-model}. However, neither of the two alternatives
provides any simplification to the optimal layout algorithm or the pricing
algorithm. For this reason, we will focus solely on
Definition~\ref{definition:separable-click-through-rates}.

\section{Layout algorithm}
\label{section:layout-algorithm}

In this section, we describe an algorithm for computing the optimal layout. In
Section~\ref{subsection:input}, we describe the input of the algorithm. In
Section~\ref{subsection:optimization-problem}, we phrase the problem of finding
the optimal layout as a discrete optimization problem. In
Section~\ref{subsection:layout-algorithm}, we describe the algorithm for finding
the optimal solution of the discrete optimization problem.

\subsection{Input}
\label{subsection:input}

The input of the algorithm consists of four parts. The first part specifies the
ad candidates, the second part specifies the structure of the grid, the third
part specifies the position multipliers, and the fourth part is the reserve
price.

The first part of the input consists of a positive integer $N$ and a list of $N$
ad candidates. Each ad candidate $i=1,2,\dots,N$ consists of the bid $b_i$,
width (single-slot or double-wide), and ad factor $\alpha_i$. We assume that
$\alpha_1, \alpha_2, \dots, \alpha_N$ are positive real numbers and that $b_1,
b_2, \dots, b_N$ are non-negative real numbers.

The second part of the input consists of positive integers $S$ and $K$ and two
lists of integers $a_1, a_2, \dots, a_K$ and $b_1, b_2, \dots, b_K$. The number
$S$ is the total number of squares in the grid. (The algorithm does not need to
know the number of rows and the number of columns of grid.) The numbers $a_1,
a_2, \dots, a_K$ and $b_1, b_2, \dots, b_K$ encode intervals $[a_1, b_1], [a_2,
b_2], \dots, [a_K, b_K]$. The squares in an interval $[a_k, b_k]$,
$k=1,2,\dots,K$, must lie within the same row of the grid and they must be
available for placing ads; see Figure~\ref{figure:blocks}.  This information is
necessary so that the algorithm can determine if a double-wide ad can be placed
at a particular position or not. Namely, two squares occupied by the same
double-wide ad must both lie in same interval $[a_k, b_k]$. We assume that $1
\le a_1 \le b_1 < a_2 \le b_2 < a_3 \le \dots < a_n \le b_n \le S$. In other
words, we assume that the intervals $[a_1, b_1], [a_2, b_2], \dots, [a_K, b_K]$
are sorted and pairwise disjoint. A square $j$ is called \emph{available} if $j
\in \bigcup_{k=1}^K [a_k, b_k]$. A square $j$ is called \emph{unavailable} if $j
\not\in \bigcup_{k=1}^K [a_k, b_k]$.

The third part of the input are position multipliers $\beta_1, \beta_2, \dots,
\beta_S$ and $\gamma_1, \gamma_2, \dots, \gamma_{S-1}$. We assume that the
position multipliers are positive real numbers and that the two sequences are
non-increasing. That is, $\beta_1 \ge \beta_2 \ge \dots \ge \beta_S$ and
$\gamma_1 \ge \gamma_2 \ge \dots \ge \gamma_{S-1}$. (See
Definition~\ref{definition:separable-click-through-rates}.)

The fourth part of the input is a per-click reserve price $r$. It is the minimum
value of a bid required for an ad to be shown. At the same time, $r$ is the
minimum amount an advertiser is charged when a user clicks on the advertiser's
ad.

\begin{figure}
\begin{center}

\begin{tikzpicture}

\node at (1,-1) [minimum size=1cm,draw,ultra thick] (1) {1};
\node at (2,-1) [minimum size=1cm,draw,ultra thick] (2) {2};
\node at (3,-1) [minimum size=1cm,draw,ultra thick] (3) {3};
\node at (4,-1) [minimum size=1cm,draw,ultra thick,fill=gray] (4) {4};
\node at (5,-1) [minimum size=1cm,draw,ultra thick,fill=gray] (5) {5};
\node at (6,-1) [minimum size=1cm,draw,ultra thick,fill=gray] (6) {6};

\node at (1,-2) [minimum size=1cm,draw,ultra thick,fill=gray] (7) {7};
\node at (2,-2) [minimum size=1cm,draw,ultra thick,fill=gray] (8) {8};
\node at (3,-2) [minimum size=1cm,draw,ultra thick,fill=gray] (9) {9};
\node at (4,-2) [minimum size=1cm,draw,ultra thick,fill=gray] (10) {10};
\node at (5,-2) [minimum size=1cm,draw,ultra thick] (11) {11};
\node at (6,-2) [minimum size=1cm,draw,ultra thick] (12) {12};

\node at (1,-3) [minimum size=1cm,draw,ultra thick] (13) {13};
\node at (2,-3) [minimum size=1cm,draw,ultra thick,fill=gray] (14) {14};
\node at (3,-3) [minimum size=1cm,draw,ultra thick,fill=gray] (15) {15};
\node at (4,-3) [minimum size=1cm,draw,ultra thick,fill=gray] (16) {16};
\node at (5,-3) [minimum size=1cm,draw,ultra thick,fill=gray] (17) {17};
\node at (6,-3) [minimum size=1cm,draw,ultra thick,fill=gray] (18) {18};

\node at (1,-4) [minimum size=1cm,draw,ultra thick,fill=gray] (19) {19};
\node at (2,-4) [minimum size=1cm,draw,ultra thick,fill=gray] (20) {20};
\node at (3,-4) [minimum size=1cm,draw,ultra thick] (21) {21};
\node at (4,-4) [minimum size=1cm,draw,ultra thick] (22) {22};
\node at (5,-4) [minimum size=1cm,draw,ultra thick] (23) {23};
\node at (6,-4) [minimum size=1cm,draw,ultra thick,fill=gray] (24) {24};

\end{tikzpicture}

\caption{\label{figure:blocks} The figure shows an example of a grid with $R=4$
rows and $C=6$ columns. Altogether, the grid has $S=RC=24$ squares. The rows are
numbered $1,2,\dots,24$ in row-major order. There are $9$ squares available for
placing the ads. These squares have white background. The remaining $15$
squares---shown with gray background---are unavailable. For example, they are
populated by other pieces of content, e.g., non-sponsored products. The
structure of white squares is encoded with four intervals $[a_1, b_1] = [1,3]$,
$[a_2, b_2]=[11,12]$, $[a_3, b_3]=[13, 13]$, $[a_4, b_4] = [21, 23]$. Each
interval corresponds to a set of consecutive white squares that lie in the same
row of the grid.}
\end{center}
\end{figure}

\subsection{Optimization problem}
\label{subsection:optimization-problem}

The problem of finding the optimal layout of ads can be formulated as a discrete
optimization problem. The goal is to compute an assignment $\pi:\{1,2,\dots,N\}
\to \{0,1,2,\dots,S\}$ of ads to positions. If $\pi(i)=j$ and $j \neq 0$, we say
that ad $i$ is \emph{assigned} to a \emph{position} $j$. If $\pi(i)=0$, we say
that ad $i$ is \emph{unassigned}. A square $j$ is called \emph{covered} by ad
$i$ if $\pi(i)=j$ or if $\pi(i)=j-1$ and ad $i$ is double-wide. If a square is
not covered by any ad and it is available, we say that it is \emph{uncovered} or
\emph{empty}. If $0 < \pi(i) < \pi(i')$, we say that ad $i$ \emph{precedes} ad
$i'$. An example of an assignment is shown in Figure~\ref{figure:assignment}.

An assignment $\pi:\{1,2,\dots,N\} \to \{0,1,2,\dots,S\}$ must satisfy
four constraints. For all $i,i' \in \{1,2,\dots,N\}$,
\begin{enumerate}
\item If $\pi(i')=\pi(i)$ then $i'=i$ or $\pi(i')=\pi(i)=0$.
\item If $\pi(i')=\pi(i) + 1$ then $\pi(i)=0$ or ad $i$ is single-slot.
\item If ad $i$ is single-slot then either $\pi(i) = 0$ or there exists $k \in \{1,2,\dots,K\}$
such that $a_k \le \pi(i) \le b_k$.
\item If ad $i$ is double-wide then either $\pi(i) = 0$ or there exists $k \in \{1,2,\dots,K\}$
such that $a_k \le \pi(i) \le b_k - 1$.
\end{enumerate}

\begin{figure}
\begin{center}

\begin{tikzpicture}

\node at (0,2) [minimum size=1cm,draw,thick] {$i$};
\node at (1,2) [minimum size=1cm,draw,thick] {$1$};
\node at (2,2) [minimum size=1cm,draw,thick] {$2$};
\node at (3,2) [minimum size=1cm,draw,thick] {$3$};
\node at (4,2) [minimum size=1cm,draw,thick] {$4$};
\node at (5,2) [minimum size=1cm,draw,thick] {$5$};
\node at (6,2) [minimum size=1cm,draw,thick] {$6$};
\node at (7,2) [minimum size=1cm,draw,thick] {$7$};

\node at (0,1) [minimum size=1cm,draw,thick] {$\pi(i)$};
\node at (1,1) [minimum size=1cm,draw,thick] {$1$};
\node at (2,1) [minimum size=1cm,draw,thick] {$13$};
\node at (3,1) [minimum size=1cm,draw,thick] {$23$};
\node at (4,1) [minimum size=1cm,draw,thick] {$2$};
\node at (5,1) [minimum size=1cm,draw,thick] {$11$};
\node at (6,1) [minimum size=1cm,draw,thick] {$21$};
\node at (7,1) [minimum size=1cm,draw,thick] {$0$};


\node at (4,-1) [minimum size=1cm,draw,ultra thick,fill=gray] (4) {4};
\node at (5,-1) [minimum size=1cm,draw,ultra thick,fill=gray] (5) {5};
\node at (6,-1) [minimum size=1cm,draw,ultra thick,fill=gray] (6) {6};

\node at (1,-1) [minimum size=1cm, minimum width=1cm,draw,ultra thick] (SS1) {1};
\node at (1,-1) [circle,minimum size=0.7cm,draw,ultra thick] {};

\node at (2.5,-1) [minimum size=1cm,minimum width=2cm,draw,ultra thick] (XL1) {4};
\node at (2.5,-1) [circle,minimum size=0.7cm,draw,ultra thick] {};

\node at (1,-2) [minimum size=1cm,draw,ultra thick,fill=gray] (7) {7};
\node at (2,-2) [minimum size=1cm,draw,ultra thick,fill=gray] (8) {8};
\node at (3,-2) [minimum size=1cm,draw,ultra thick,fill=gray] (9) {9};
\node at (4,-2) [minimum size=1cm,draw,ultra thick,fill=gray] (10) {10};

\node at (5.5,-2) [minimum size=1cm, minimum width=2cm,draw,ultra thick] (XL2) {5};
\node at (5.5,-2) [circle,minimum size=0.7cm, draw,ultra thick] {};

\node at (2,-3) [minimum size=1cm,draw,ultra thick,fill=gray] (14) {14};
\node at (3,-3) [minimum size=1cm,draw,ultra thick,fill=gray] (15) {15};
\node at (4,-3) [minimum size=1cm,draw,ultra thick,fill=gray] (16) {16};
\node at (5,-3) [minimum size=1cm,draw,ultra thick,fill=gray] (17) {17};
\node at (6,-3) [minimum size=1cm,draw,ultra thick,fill=gray] (18) {18};

\node at (1,-3) [minimum size=1cm, minimum width=1cm,draw,ultra thick] (SS2) {2};
\node at (1,-3) [circle, minimum size=0.7cm, draw,ultra thick] {};

\node at (1,-4) [minimum size=1cm,draw,ultra thick,fill=gray] (19) {19};
\node at (2,-4) [minimum size=1cm,draw,ultra thick,fill=gray] (20) {20};
\node at (6,-4) [minimum size=1cm,draw,ultra thick,fill=gray] (24) {24};

\node at (3.5,-4) [minimum size=1cm, minimum width=2cm,draw,ultra thick] (XL3) {6};
\node at (3.5,-4) [circle, minimum size=0.7cm, draw,ultra thick] {};

\node at (5,-4) [minimum size=1cm, minimum width=1cm,draw,ultra thick] (SS3) {3};
\node at (5,-4) [circle, minimum size=0.7cm, draw,ultra thick] {};

\end{tikzpicture}

\caption{\label{figure:assignment} The table at the top shows an assignment
$\pi:\{1,2,\dots,N\} \to \{0,1,2,\dots,S\}$ of $N=7$ ads to a grid with $S=24$
squares. Out of the $7$ ads, $N_1=3$ are single-slot and $N_2=4$ are
double-wide. The single-slot ads are numbered $1,2,3$ and the double-wide are
numbered $4,5,6,7$. Ads $1,2,3,4,5,6$ are assigned to a position, ad $7$ is
unassigned. The grid together with the assigned ads is shown at the bottom. The
grid has $9$ squares available for placing ads, shown with white background.
These squares have numbers $1,2,3,11,12,13,21,22,23$. All available squares are
covered by an ad. The ad numbers are shown in circles.}
\end{center}
\end{figure}

The goal is to find an assignment $\pi$ that satisfies the four constraints
above and maximizes the expected sum of bids of ads the user clicks on. This
number is called \emph{efficiency} or \emph{first-price revenue}. It
is defined as
\begin{equation}
\label{equation:efficiency}
\sum_{i~:~\pi(i) \neq 0} b_i \PCTR_{i,\pi(i)}
= \sum_{\substack{\text{single-slot ad $i$} \\ \pi(i) \neq 0}} b_i \alpha_i \beta_{\pi(i)}
+ \sum_{\substack{\text{double-wide ad $i$} \\ \pi(i) \neq 0}} b_i \alpha_i \gamma_{\pi(i)} \: .
\end{equation}
In other words, efficiency expressed in~\eqref{equation:efficiency} is the
objective function of the optimization problem. The equality in
\eqref{equation:efficiency} follows from equation
\eqref{equation:separable-model} in
Definition~\ref{definition:separable-click-through-rates}.

\subsection{Algorithm}
\label{subsection:layout-algorithm}

In this section we describe an algorithm that finds the optimal assignment
$\pi$. The algorithm consists of two steps: preprocessing and dynamic
programming.

\subsubsection{Preprocessing}
\label{subsubsection:preprocessing}

Preprocessing consists of three steps. The time complexity of all three steps is
$O(N \log N)$. The additional memory complexity is constant.

As the first step, all ads with bids below $r$ are removed. After this
step $b_i \ge r$ for all $i=1,2,\dots,N$.

As the second step, the algorithm adjusts the number of single-slot ads. Let
$T=\sum_{k=1}^K (b_k - a_k + 1)$ be the total number of available squares. If
necessary, the algorithm adds fictitious single-slot ads so that their number is
at least $T$. Namely, if the number of single-slot ads is below $T$, the
algorithm creates fictitious single-slot ads with bid $b_i=0$ and an arbitrary
ad factor $\alpha_i$, say, $\alpha_i=1$. The purpose of the fictitious ads is to
ensure that there are no gaps in the optimal assignment.

As the third step, the algorithm sorts the ads according to their width in
increasing order. Ads with same width are sorted according to $b_i \alpha_i$ in
decreasing order. Let $N_1, N_2$ be the number of single-slot and double-wide
ads respectively. After sorting, ads $1,2,\dots,N_1$ are single-slot, ads
$N_1+1, N_1+2, \dots, N_1+N_2$ are double-wide, and
\begin{align}
\label{equation:single-slot-ads-sorted}
b_1 \alpha_1 & \ge b_2 \alpha_2 \ge \dots \ge b_{N_1} \alpha_{N_1} \: , \\
\label{equation:double-wide-ads-sorted}
b_{N_1+1} \alpha_{N_1+1} & \ge b_{N_1+2} \alpha_{N_1+2} \ge \dots \ge b_{N_1 + N_2} \alpha_{N_1 + N_2} \: .
\end{align}

It easy to prove that there exist integers $N_1'$ and $N_2'$ and an optimal
assignment $\pi$ such that $0 \le N_1' \le N_1$, $0 \le N_2' \le N_2$ and
\begin{equation}
\label{equation:prefix-conditions}
\begin{gathered}
0 < \pi(1) < \pi(2) < \dots < \pi(N_1') \: , \\
\pi(N_1'+1) = \pi(N_1'+2) = \dots = \pi(N_1) = 0 \: , \\
0 < \pi(N_1+1) < \pi(N_1+2) < \dots < \pi(N_1+N_2') \: , \\
\pi(N_1+N_2'+1) = \pi(N_1+N_2'+2) = \dots = \pi(N_1+N_2) = 0 \: .
\end{gathered}
\end{equation}
Conditions in \eqref{equation:prefix-conditions} follow from the well-known
rearrangement inequality stated as Lemma~\ref{lemma:rearrangement-inequality}
below and from assumptions $\beta_1 \ge \beta_2 \ge \dots \ge \beta_S$ and
$\gamma_1 \ge \gamma_2 \ge \dots \ge \gamma_{S-1}$ and conditions
\eqref{equation:single-slot-ads-sorted} and
\eqref{equation:double-wide-ads-sorted}.
(Lemma~\ref{lemma:rearrangement-inequality} can be proved by a simple exchange
argument.)

\begin{lemma}[Rearrangement inequality]
\label{lemma:rearrangement-inequality}
Let $x_1 \ge x_2 \ge \dots \ge x_n$ and $y_1 \ge y_2 \ge \dots \ge y_n$ be real
numbers. Then, for any permutation $\sigma:\{1,2,\dots,N\} \to \{1,2,\dots,N\}$,
$$
x_1 y_1 + x_2 y_2 + \dots + x_n y_n \ge
x_1 y_{\sigma(1)} + x_2 y_{\sigma(2)} + \dots + x_n y_{\sigma(n)} \: .
$$
\end{lemma}

The search for the optimal assignment can be limited to assignments $\pi$ that
satisfy \eqref{equation:prefix-conditions}. Furthermore, the search can be
limited to assignments that do not have any \emph{gaps}. That is, there cannot
exist an empty square $j$ and an ad assigned to a position $j' > j$. This
follows from the fact that the number of single-slot ads is at least $T$. Thus,
there exist an optimal assignment that does not have any gaps and at the same
satisfies \eqref{equation:prefix-conditions}.

\subsubsection{Dynamic programming}
\label{subsubsection:dynamic-programming}

\begin{figure}
\begin{center}
\begin{tikzpicture}

\begin{scope}[every node/.style={circle,ultra thick,draw}]
\node (N00) at (0,6) {$0, 0$};
\node (N01) at (0,3) {$0, 1$};
\node (N02) at (0,0) {$0, 2$};

\node (N10) at (3,6) {$1, 0$};
\node (N11) at (3,3) {$1, 1$};
\node (N12) at (3,0) {$1, 2$};

\node (N20) at (6,6) {$2, 0$};
\node (N21) at (6,3) {$2, 1$};
\node (N22) at (6,0) {$2, 2$};

\node (N30) at (9,6) {$3, 0$};
\node (N31) at (9,3) {$3, 1$};
\node (N32) at (9,0) {$3, 2$};

\node (N40) at (12,6) {$4, 0$};
\node (N41) at (12,3) {$4, 1$};
\node (N42) at (12,0) {$4, 2$};

\node (N50) at (15,6) {$5, 0$};
\node (N51) at (15,3) {$5, 1$};
\node (N52) at (15,0) {$5, 2$};
\end{scope}

\begin{scope}[>={Latex[length=4mm]},ultra thick]

\path [->] (N00) edge (N01);
\path [->] (N01) edge (N02);

\path [->] (N10) edge (N11);
\path [->] (N11) edge (N12);

\path [->] (N20) edge (N21);
\path [->] (N21) edge (N22);

\path [->] (N30) edge (N31);
\path [->] (N31) edge (N32);

\path [->] (N40) edge (N41);
\path [->] (N41) edge (N42);

\path [->] (N50) edge (N51);
\path [->] (N51) edge (N52);

\path [->] (N00) edge (N10);
\path [->] (N10) edge (N20);
\path [->] (N20) edge (N30);
\path [->] (N30) edge (N40);
\path [->] (N40) edge (N50);

\path [->] (N01) edge (N11);
\path [->] (N11) edge (N21);
\path [->] (N21) edge (N31);
\path [->] (N31) edge (N41);
\path [->] (N41) edge (N51);

\path [->] (N02) edge (N12);
\path [->] (N12) edge (N22);
\path [->] (N22) edge (N32);
\path [->] (N32) edge (N42);
\path [->] (N42) edge (N52);

\end{scope}

\end{tikzpicture}
\caption{\label{figure:grid-graph} The figure shows an example of a grid graph.
The graph has $3 \times 6 = 18$ vertices and $2 \times 6 + 3 \times 5 = 27$
directed edges. The vertices of the graph are pairs $(u,v)$ where $u \in
\{0,1,2\}$ and $v \in \{0,1,2,3,4,5\}$. There is a directed edge from vertex
$(u,v)$ to vertex $(u+1, v)$. Similarly, there is a directed edge from vertex
$(u,v)$ to vertex $(u, v+1)$. Note that the graph has no directed cycles.}
\end{center}
\end{figure}
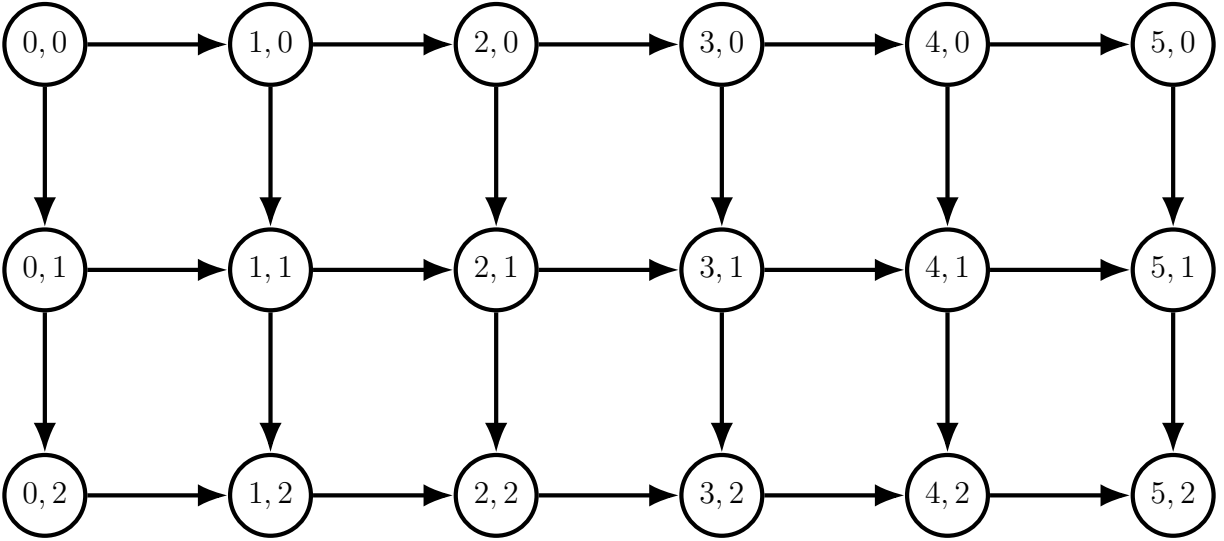

The problem of finding optimal assignment $\pi$ without gaps satisfying
\eqref{equation:prefix-conditions} can be reduced to the problem of finding the
path of maximum weight in a directed acyclic graph $G$ with edge weights.
The graph $G$ is a subgraph of a two-dimensional grid graph; see
Figure~\ref{figure:grid-graph}.

Before we specify the vertices and the edges of $G$, recall that $T$ is the
number of available squares. Let $s_1 < s_2 < \dots < s_T$ be the available
squares sorted in increasing order. Recall that that $N_1$ and $N_2$ are the
numbers of single-slot and double-wide ads respectively. Finally, recall that
$N_1 \ge T$.

The set of vertices of graph $G$ is
\begin{equation}
\label{equation:vertex-set}
V = \{(u,v) ~:~ u, v \in \Z, 0 \le u \le N_1, 0 \le v \le N_2, u + 2v \le T \} \: .
\end{equation}
Intuitively, a vertex $(u,v) \in V$ corresponds to the set of all assignments
$\pi$ that do not have any gaps, satisfy \eqref{equation:prefix-conditions}, and
in which (single-slot) ads $1,2,\dots,u$ and (double-wide) ads
$N_1+1,N_1+2,\dots,N_1+v$ are assigned to some position and all other ads are
unassigned. Note that any such assignment covers the first $u+2v$ available
squares, i.e., squares $s_1, s_2, \dots, s_{u+2v}$.

\begin{figure}
\begin{center}

\begin{subfigure}[b]{\linewidth}
\centering
\begin{tikzpicture}

\node at (1,-1) [minimum size=1cm,draw,ultra thick] (1) {1};
\node at (2,-1) [minimum size=1cm,draw,ultra thick,fill=gray] (2) {2};
\node at (3,-1) [minimum size=1cm,draw,ultra thick,fill=gray] (3) {3};
\node at (4,-1) [minimum size=1cm,draw,ultra thick,fill=gray] (4) {4};

\node at (1,-2) [minimum size=1cm,draw,ultra thick,fill=gray] (5) {5};
\node at (2,-2) [minimum size=1cm,draw,ultra thick,fill=gray] (6) {6};
\node at (3,-2) [minimum size=1cm,draw,ultra thick,fill=gray] (7) {7};
\node at (4,-2) [minimum size=1cm,draw,ultra thick,fill=gray] (8) {8};

\node at (1,-3) [minimum size=1cm,draw,ultra thick,fill=gray] (9) {9};
\node at (2,-3) [minimum size=1cm,draw,ultra thick,fill=gray] (10) {10};
\node at (3,-3) [minimum size=1cm,draw,ultra thick] (11) {11};
\node at (4,-3) [minimum size=1cm,draw,ultra thick] (12) {12};

\node at (1,-4) [minimum size=1cm,draw,ultra thick,fill=gray] (13) {13};
\node at (2,-4) [minimum size=1cm,draw,ultra thick,fill=gray] (14) {14};
\node at (3,-4) [minimum size=1cm,draw,ultra thick,fill=gray] (15) {15};
\node at (4,-4) [minimum size=1cm,draw,ultra thick,fill=gray] (16) {16};

\node at (1,-5) [minimum size=1cm,draw,ultra thick,fill=gray] (17) {17};
\node at (2,-5) [minimum size=1cm,draw,ultra thick] (18) {18};
\node at (3,-5) [minimum size=1cm,draw,ultra thick] (19) {19};
\node at (4,-5) [minimum size=1cm,draw,ultra thick,fill=gray] (20) {20};

\end{tikzpicture}

\caption{Grid.\label{subfigure:grid}}
\end{subfigure}

\vspace{2cm}

\begin{subfigure}[b]{\linewidth}
\centering
\begin{tikzpicture}

\begin{scope}[every node/.style={circle,ultra thick,draw}]
\node (N00) at (0,6) {$0, 0$};
\node[dashed] (N01) at (0,3) {$0, 1$};
\node[dashed] (N02) at (0,0) {$0, 2$};

\node (N10) at (3,6) {$1, 0$};
\node (N11) at (3,3) {$1, 1$};
\node (N12) at (3,0) {$1, 2$};

\node (N20) at (6,6) {$2, 0$};
\node (N21) at (6,3) {$2, 1$};

\node (N30) at (9,6) {$3, 0$};
\node (N31) at (9,3) {$3, 1$};

\node (N40) at (12,6) {$4, 0$};

\node (N50) at (15,6) {$5, 0$};

\end{scope}

\begin{scope}[>={Latex[length=4mm]},ultra thick]

\path [->, dashed] (N01) edge node[right]{$b_7 \alpha_7 \gamma_{12}$} (N02);

\path [->] (N10) edge node[right]{$b_6 \alpha_6 \gamma_{11}$} (N11);
\path [->] (N11) edge node[right]{$b_7 \alpha_7 \gamma_{18}$} (N12);

\path [->] (N30) edge node[right]{$b_6 \alpha_6 \gamma_{18}$} (N31);

\path [->] (N00) edge node[above]{$b_1 \alpha_1 \beta_1$} (N10);
\path [->] (N10) edge node[above]{$b_2 \alpha_2 \beta_{11}$} (N20);
\path [->] (N20) edge node[above]{$b_3 \alpha_3 \beta_{12}$} (N30);
\path [->] (N30) edge node[above]{$b_4 \alpha_4 \beta_{18}$} (N40);
\path [->] (N40) edge node[above]{$b_5 \alpha_5 \beta_{19}$} (N50);

\path [->,dashed] (N01) edge node[above]{$b_1 \alpha_1 \beta_{12}$} (N11);
\path [->] (N11) edge node[above]{$b_2 \alpha_2 \beta_{18}$} (N21);
\path [->] (N21) edge node[above]{$b_3 \alpha_3 \beta_{19}$} (N31);

\path [->,dashed] (N02) edge node[above]{$b_1 \alpha_1 \beta_{19}$} (N12);

\end{scope}
\end{tikzpicture}
\caption{Graph $G$.\label{subfigure:graph}}
\end{subfigure}

\caption{\label{figure:graph} Figure~\ref{subfigure:grid} shows an example of a
grid with $S=20$ squares. The grid has $T=5$ available squares. The available
squares are $s_1 = 1, s_2 = 11, s_3 = 12, s_4 = 18, s_5 = 19$ and they are
depicted with white background. The remaining $15$ squares are unavailable and
they are depicted with gray background. The structure of the available squares
is encoded as three intervals $[a_1, b_1] = [1,1]$, $[a_2, b_2] = [11,12]$,
$[a_3, b_3] = [18,19]$. Figure~\ref{subfigure:graph} shows the graph $G$
corresponding to the grid and $N=7$ ads. Out of the $7$ ads, $N_1=5$ are
single-slot and $N_2=2$ are double-wide. Ads $1,2,3,4,5$ are single-slot and ads
$6,7$ are double-wide. Vertices and edges that are not reachable from vertex
$(0,0)$ are shown with dashed lines.}
\end{center}
\end{figure}
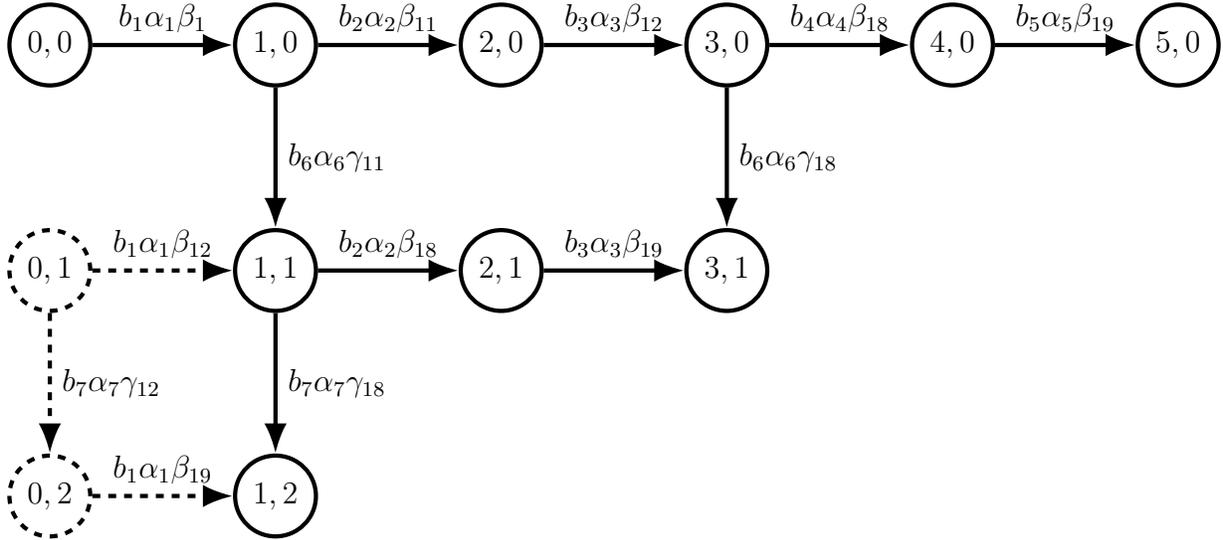

The set of edges of $G$ and their weights is a defined as follows. If $(u-1,v),
(u, v) \in V$ then there exists a directed edge from $(u-1,v)$ to $(u, v)$
with weight
$$
w((u-1,v), (u,v)) = b_u \alpha_u \beta_{s_{u + 2v}} \: .
$$
This edge corresponds to placing (single-slot) ad $u$ at position
$s_{u+2v}$. If $(u,v-1), (u, v) \in V$ and $s_{u+2v} = s_{u+2v-1} + 1$ then
there exists an edge from $(u,v-1)$ to $(u, v)$ with weight
$$
w((u,v-1), (u,v)) = b_{N_1 + v} \alpha_{N_1 + v} \gamma_{s_{u + 2v - 1}} \: .
$$
This edge corresponds to placing (double-wide) ad $N_1+v+1$ at position
$s_{u+2v-1}$. That is, the ad covers two adjacent available squares $s_{u+2v-1}$
and $s_{u+2v}$. For convenience, if there is no edge from a vertex $(u,v)$ to
a vertex $(u',v')$, we define the weight $w((u,v), (u',v')) = -\infty$. An
example of graph $G$ is shown in Figure~\ref{figure:graph}.

An assignment $\pi$ without gaps satisfying \eqref{equation:prefix-conditions}
corresponds to a directed path in $G$ starting from vertex $(0,0)$. The expected
first-price revenue (efficiency) of the assignment is the sum of the weights of
the edges on the path. Thus, the optimal assignment corresponds to the path of
the maximum weight that starts from vertex $(0,0)$.

The path of the maximum weight can be found by dynamic programming. The
algorithm enumerates the vertices in row-major order (or any other topological
order). For each vertex $(u,v) \in V$, it computes and stores the maximum weight
$M[u,v]$ of a path starting at vertex $(0,0)$ and ending at vertex $(u,v)$. The
value $M[u,v]$ is computed from the values $M[u-1,v]$ and $M[u,v-1]$ computed
previously and the edge weights $w((u-1,v), (u,v))$ and $w((u,v-1), (u,v))$
using the formula
$$
M[u,v] =
\begin{cases}
0 & \text{if $u=v=0$,} \\
M[u-1,v] + w((u-1,v), (u,v))  & \text{if $u > 0$ and $v = 0$,} \\
M[u,v-1] + w((u,v-1), (u,v))  & \text{if $u = 0$ and $v > 0$,} \\
\begin{aligned}
\max \big\{ & M[u-1,v] + w((u-1,v), (u,v)), \\
        & M[u,v-1] + w((u,v-1), (u,v)) \big\}
\end{aligned}
& \text{if $u > 0$ and $v > 0$.}
\end{cases}
$$
For each vertex $(u,v) \in V$ such that $u > 0$ and $v > 0$, the algorithm
computes and stores the record of which of the two branches in the maximum in
the formula for $M[u,v]$ was taken. In order to compute the path, the algorithm
finds the vertex $(u^*,v^*)$ with the maximum value $M[u^*,v^*]$. Starting from
$(u^*, v^*)$, it traverses the edges in the reverse direction until it reaches
vertex $(0,0)$. If there are two edges to choose from---which can happen only if
$u > 0$ and $v > 0$---the algorithm chooses the edge corresponding to the
recorded choice made earlier when $M[u,v]$ was computed.

The time and space complexity of the algorithm is proportional to the number of
vertices of the graph. The number of vertices is at most
$$
(\min\{N_1, T\} + 1)(\min\{N_2, \lfloor T/2 \rfloor\} + 1) \le (T+1)(N_2+1) \le (N_1 + 1)(N_2 + 1) \: .
$$

\section{Pricing algorithm}
\label{section:pricing-algorithm}

In this section, we describe the pricing algorithm. The algorithm computes the
price per click for each ad chosen by the layout algorithm described in
Section~\ref{section:layout-algorithm}. The algorithm can be thought of as a
generalization of pricing rule in generalized second-price (GSP) auction.

The algorithm described in this section implicitly assumes that each ad comes
from a different advertiser. This is a standard assumption made (implicitly or
explicitly) by the majority of ad auctions, including GSP and VCG. On Instacart,
this assumption is violated. In Section~\ref{section:extensions}, we describe an
extension of the pricing algorithm for the setting where multiple ads from the
same advertiser participate in the auction.

The pricing algorithm continues where the layout algorithm left off. In
particular, the inputs to the pricing algorithm are identical. We assume that
preprocessing, as described in Section~\ref{subsubsection:preprocessing}, has
been done. We assume that the graph $G$, the array $M$, the optimal assignment
$\pi^*$ have been computed by the layout algorithm, as described in
Section~\ref{subsubsection:dynamic-programming}.

The price per click $p_i$ of an ad $i$ is computed as the smallest value of the
bid $b_i$ such that the optimal ranking $\pi^*$ computed by the layout algorithm
remains the same. In this sense, the pricing algorithm is a generalization of
GSP auction. More specifically, the price is computed as the maximum of three
values,
\begin{equation}
\label{equation:maximum-of-three-values}
p_i = \max\{r, c_i, d_i\} \: .
\end{equation}

The first value in \eqref{equation:maximum-of-three-values} is the reserve price
$r$. Clearly, if the value of $b_i$ decreases below $r$, the ad $i$ is no longer
assigned to any position in the optimal ranking.

The second value in \eqref{equation:maximum-of-three-values}, $c_i$, depends on
the next ad of the same width. That is, assume that ad $i$ is a single-slot and
ad $i+1$ is the next single-slot ad in the ordering
\eqref{equation:single-slot-ads-sorted}. Ad $i$ swaps place with ad $i+1$ when
the value of $b_i$ goes below
\begin{equation}
\label{equation:standard-gsp-price}
c_i = b_{i+1} \alpha_{i+1} / \alpha_i \: .
\end{equation}
For $i=N_1$, we define $c_i = -\infty$, since ad $i+1$ either does not exist or
it is double-wide. If ad $i$ is double-wide, we define $c_i$ using the formula
\eqref{equation:standard-gsp-price} provided that $i < N_1+N_2$. For
$i=N_1+N_2$, we define $c_i = -\infty$, since ad $i+1$ does not exist. The
formula \eqref{equation:standard-gsp-price} is the price for the standard GSP
auction.

The third value in \eqref{equation:maximum-of-three-values}, $d_i$, depends on
ads of different widths. It can be computed by considering alternative rankings
in which ad $i$ is moved after ads of different width. In order to compute
$d_i$, the algorithm needs the value $M[u,v]$ and an additional value $L[u,v]$
for each vertex $(u,v)$ of the graph $G$. The value $L[u,v]$ is defined as the
maximum weight of a directed path in $G$ starting in vertex $(u,v)$. The numbers
$L[u,v]$ can be computed by dynamic programming. The algorithm enumerates the
vertices in $V$ in reversed row-major order (or any other reversed topological
order). The value $L[u,v]$ can be computed from the values of $L[u+1,v]$ and
$L[u,v+1]$ computed previously and the edge weights $w((u,v), (u+1,v))$ and
$w((u,v), (u,v+1))$ using the formula
$$
L[u,v] =
\begin{cases}
0 & \text{if $(u+1, v) \not \in V$ and $(u, v+1) \not \in V$,} \\
w((u,v), (u+1,v)) + L[u+1,v]  & \text{if $(u+1, v) \in V$ and $(u, v+1) \not \in V$,} \\
w((u,v), (u,v+1)) + L[u,v+1]  & \text{if $(u+1, v) \not \in V$ and $(u, v+1) \in V$,} \\
\begin{aligned}
\max \big\{ & w((u,v), (u+1,v)) + L[u+1,v], \\
        & w((u,v), (u,v+1)) + L[u,v+1] \big\}
\end{aligned}
& \text{if $(u+1, v) \in V$ and $(u, v+1) \in V$.}
\end{cases}
$$
Using $L$ we can compute the value $d_i$. In order to compute it, we consider
two cases based on the width of ad $i$.

\textbf{Case 1: Ad $i$ is single-slot.} Let $v = \{ i' ~:~ N_1+1 \le i' \le N_1
+ N_2, 0 < \pi^*(i') < \pi^*(i) \}$ be the number of double-wide ads in the
optimal assignment $\pi^*$ preceding ad $i$. Then, the objective value of
$\pi^*$ is
\begin{equation}
\label{equation:optimal-assignment-decomposition-single-slot}
M[i-1,v] + w((i-1,v), (i,v)) + L[i,v] = M[i-1,v] + b_i \alpha_i \beta_{s_{i+2v}} + L[i,v] \: .
\end{equation}
The expression above is a linear function of the bid $b_i$. For any $v' \in
\{0,1,\dots,N_2\}$ such that $((i-1,v'), (i,v')) \in E$, consider the
alternative assignment with $v'$ double-wide ads preceding ad $i$ and
objective value
\begin{equation}
\label{equation:alternative-assignment-decomposition-single-slot}
M[i-1,v'] + w((i-1,v'), (i,v')) + L[i,v'] = M[i-1,v'] + b_i \alpha_i \beta_{s_{i+2v'}} + L[i,v'] \: .
\end{equation}
The objective value of the alternative assignment is a linear function of $b_i$.
Since $\pi^*$ is optimal, for all $v'=0,1,\dots,N_2$ such that $((i-1,v'),
(i,v')) \in E$,
\begin{equation}
\label{equation:linear-inequalities-single-slot}
M[i-1,v] + b_i \alpha_i \beta_{s_{i+2v}} + L[i,v] \ge M[i-1,v'] + b_i \alpha_i \beta_{s_{i+2v'}} + L[i,v'] \: .
\end{equation}
These are linear inequalities in $b_i$ indexed by $v'$. We define $d_i$ as the
smallest value of $b_i$ such that all the inequalities remain valid. The
explicit formula is
\begin{multline}
\label{equation:critical-bid-single-slot}
d_i = \max \bigg\{ \frac{M[i-1,v'] + L[i,v'] - M[i-1,v] - L[i,v]}{\alpha_i (\beta_{s_{i+2v}} - \beta_{s_{i+2v'}})} \\
~:~ ((i-1,v'), (i,v')) \in E, \ \beta_{s_{i+2v}} > \beta_{s_{i+2v'}} \bigg\} \: .
\end{multline}

\textbf{Case 2: Ad $i$ is double-wide.} Let $u = \{ i' ~:~ 1 \le i' \le N_1, 0 <
\pi^*(i') < \pi^*(i) \}$ be the number of single-slot ads in the optimal
assignment $\pi^*$ preceding ad $i$. Then, the objective value of $\pi^*$ is
\begin{equation}
\label{equation:optimal-assignment-decomposition-double-wide}
M[u,i-1] + w((u,i-1), (u,i)) + L[u,i] = M[u,i-1] + b_i \alpha_i \gamma_{s_{u+2i}} + L[u,i] \: .
\end{equation}
The expression above is a linear function of the bid $b_i$. For any $u' \in
\{0,1,\dots,N_2\}$ such that $((u',i-1), (u',i)) \in E$, consider the
alternative assignment with $u'$ single-slot ads preceding ad $i$ and
objective value
\begin{equation}
\label{equation:alternative-assignment-decomposition-double-wide}
M[u',i-1] + w((u',i-1), (u',i)) + L[u',i] = M[u',i-1] + b_i \alpha_i \gamma_{s_{u'+2i}} + L[u',i]
\end{equation}
The objective value of the alternative assignment is a linear function of $b_i$.
Since $\pi^*$ is optimal, for all $u'=0,1,\dots,N_1$ such that $((u',
i-1), (u', i)) \in E$,
\begin{equation}
\label{equation:linear-inequalities-double-wide}
M[u,i-1] + b_i \alpha_i \gamma_{s_{u+2i}} + L[u,i] \ge M[u',i-1] + b_i \alpha_i \gamma_{s_{u'+2i}} + L[u',i] \: .
\end{equation}
These are linear inequalities in $b_i$ indexed by $u'$. We define $d_i$ as the
smallest value of $b_i$ such that all the inequalities remain valid. The
explicit formula is
\begin{multline}
\label{equation:critical-bid-double-wide}
d_i = \max \bigg\{ \frac{M[u',i-1] + L[u',i] - M[u,i-1] - L[u,i]}{\alpha_i (\gamma_{s_{u+2i}} - \gamma_{s_{u'+2i}})} \\
~:~ ((u', i-1), (u', i)) \in E, \ \gamma_{s_{u+2i}} > \gamma_{s_{u'+2i}} \bigg\} \: .
\end{multline}

The time and space complexity needed to compute $L$ is proportional to the
number of vertices of the graph $G$ which is upper bounded by $(N_1+1)(N_2+1)$.
The right-hand side of \eqref{equation:critical-bid-single-slot} can be
evaluated in time $O(N_2)$ by trying all possible values of $v'=0,1,\dots,N_2$.
Thus, the value $d_i$ for all single-slot ads $i=1,2,\dots,N_1$ can be computed
in $O(N_1N_2)$ time. The right-hand side of
\eqref{equation:critical-bid-double-wide} can be evaluated in time $O(N_1)$ by
trying all possible values of $u'=0,1,\dots,N_1$. Thus, the values $d_i$ for all
double-wide ads $i=N_1+1,N_1+2,\dots,N_1+N_2$ can be computed in $O(N_1N_2)$
time. Altogether, excluding preprocessing, the time and space complexity of the
pricing algorithm is $O(N_1N_2)$. In particular, the complexity is the same as
the complexity of the layout algorithm.

\subsection{VCG pricing}
\label{subsection:vcg-pricing}

An alternative pricing scheme can be obtained by application of
Vickrey-Clarke-Groves (VCG) mechanism~\citep[Chapter
9.3.3]{Nisan-Roughgarden-Tardos-Vazirani-2007}. The price for ad $i$ is computed
from the so called \emph{externality} that ad $i$ places on the other ads.
Externality of ad $i$ depends on the optimal assignment $\pi^*$ and an
alternative assignment $\pi_i^*$. The alternative assignment $\pi_i^*$ is the
optimal assignment for a modified input. The modified input is identical to the
original input with the exception that bid $b_i$ is set to $0$. Externality of
ad $i$ is defined as
\begin{equation}
\label{equation:externality}
e_i
= \left( \sum_{\substack{j~:~\pi_i^*(j) \neq 0 \\ j \neq i}} b_j \PCTR_{j,\pi_i^*(j)} \right) - \left( \sum_{\substack{j~:~\pi^*(j) \neq 0 \\ j \neq i}} b_j \PCTR_{j,\pi^*(j)} \right) \: .
\end{equation}
We can express the externality explicitly using formula
\eqref{equation:separable-model} for the separable click-through rate model. In
order to simplify the notation, we adopt the convention that $\beta_0 = \gamma_0
= 0$.
\allowdisplaybreaks
\begin{align*}
e_i
& = \left(\sum_{\substack{\text{single-slot ad $j$} \\ j \neq i}} b_j \alpha_j \beta_{\pi_i^*(j)}
+ \sum_{\substack{\text{double-wide ad $j$} \\ j \neq i}} b_j \alpha_j \gamma_{\pi_i^*(j)} \right) \\
& \qquad - \left( \sum_{\substack{\text{single-slot ad $j$} \\ j \neq i}} b_j \alpha_j \beta_{\pi^*(j)}
 + \sum_{\substack{\text{double-wide ad $j$} \\ j \neq i}} b_j \alpha_j \gamma_{\pi^*(j)} \right) \\
& = \sum_{\substack{\text{single-slot ad $j$} \\ j \neq i}} b_j \alpha_j \left( \beta_{\pi_i^*(j)} - \beta_{\pi^*(j)} \right)
 \quad + \quad \sum_{\substack{\text{double-wide ad $j$} \\ j \neq i}} b_j \alpha_j \left( \gamma_{\pi_i^*(j)} - \gamma_{\pi^*(j)} \right) \: .
\end{align*}
The VCG price for ad $i$ is computed from $e_i$ by dividing it by the
click-through rate. Formally,
$$
p_i
= \frac{e_i}{\PCTR_{i,\pi^*(i)}}
=
\begin{cases}
\dfrac{e_i}{\alpha_i \beta_{\pi^*(i)}} & \text{if ad $i$ is single-slot,} \\[0.7cm]
\dfrac{e_i}{\alpha_i \gamma_{\pi^*(i)}} & \text{if ad $i$ is double-wide.}
\end{cases}
$$

It is worth noting that the optimality of $\pi^*$ and $\pi_i^*$ for their
respective inputs implies the inequalities
$$
0 \le e_i \le b_i {\PCTR_{i,\pi^*(i)}} \: .
$$
These inequalities imply that
$$
0 \le p_i \le b_i \: .
$$
In other words, the per-click price of ad $i$ is always between $0$ and the
per-click bid $b_i$.

It is well-known that VCG prices make the auction mechanism truthful. That is,
assuming each advertiser has a quasi-linear utility function, the dominant
strategy of each advertiser is to bid their true value. For details,
see~\citet[Chapter 9.3.3]{Nisan-Roughgarden-Tardos-Vazirani-2007}.

However, compared to GSP prices, VCG prices have a computational disadvantage.
They require computation of $N$ alternative optimal assignments $\pi_1^*,
\pi_2^*, \dots, \pi_N^*$. Each alternative assignment takes $O(N_1 N_2)$ time to
compute. This brings the overall time complexity of computing VCG prices to
$O(NN_1N_2)$, which is a multiplicative factor $N$ worse than the time
complexity of the algorithm for computing GSP prices.

\section{Extensions}
\label{section:extensions}

In Sections~\ref{subsection:single-slot-vs-double-wide} and
\ref{subsection:multiple-ads-from-the-same-advertiser}, we describe two
extensions of the algorithm. In both of these extensions, the algorithms receive
an additional input that specifies the advertiser (advertiser identifier) for
each ad $i=1,2,\dots,N$. We denote by $f_i$ the advertiser\footnote{An
advertiser can be thought of as an agent (player) in the game-theoretic
mechanism implemented by the auction. In this paper, however, we do not analyze
the equilibria of the game or optimal strategies for the players for either the
algorithms described in Section~\ref{section:layout-algorithm} and
\ref{section:pricing-algorithm} or the two extensions. As far as we know, these
problems are widely open.} for ad $i$ and for concreteness we assume that $f_i
\in \N$. If $f_i=f_j$, we say ads $i$ and $j$ have the same the same advertiser.

\subsection{Single-slot vs. double-wide}
\label{subsection:single-slot-vs-double-wide}

We extend the layout and pricing algorithm to the case when each advertiser
submits to the auction at most one single-slot ad and at most one double-wide
ad, each with its own bid, and the auction has to choose at most one ad from
each advertiser.

Suppose there are $K$ advertisers with two ads (one single-slot ad and one
double-wide ad). There are $2^K$ combinations of widths of the ads for the $K$
advertisers. In order to compute the optimal layout, the algorithm tries all
$2^K$ combinations. For each combination $C$, for each of the $K$ advertisers
the algorithm selects the ad with the width according to $C$ (and removes the ad
of the other width), it runs the algorithm described in
Section~\ref{section:layout-algorithm} with the selected ads as well as ads from
advertisers with only one ad, and it computes the optimal assignment
corresponding to $C$, which we denote $\pi_C^*$. Out of the $2^K$ assignments,
the algorithm selects the assignment with the largest objective value. Let $C^*$
be the combination corresponding to the assignment with the largest objective
value. In order to compute the per-click prices, for each of the $K$ advertisers
the algorithm selects the ad the width according to $C^*$ (and removes the other
ad of the other width) and it runs the algorithm described in
Section~\ref{section:pricing-algorithm} with the selected ads as well ads from
advertisers with only one ad.

\subsection{Many ads from the same advertiser}
\label{subsection:multiple-ads-from-the-same-advertiser}

We extend the layout and pricing algorithm to the case when an advertiser can
submit any number of single-slot ads and any number of double-wide ads, each
with their own bid, and the auction can choose to show any subset of the ads,
potentially multiple ads from the same advertiser.

The optimal assignment $\pi^*$ is computed using the layout algorithm from
Section~\ref{section:layout-algorithm} without any change. In particular,
$\pi^*$ does not depend on the identity of the advertisers. However, the pricing
algorithm is modified substantially.

The pricing algorithm computes prices for each advertiser separately. Let $f \in
\N$ be an advertiser. Suppose ads $i_1, i_2, \dots, i_K$ are all the ads from
advertiser $f$ that $\pi^*$ assigns to some position. Suppose that the ads are
sorted according to the their position in the optimal assignment. That is,
$$
0 < \pi^*(i_1) < \pi^*(i_2) < \dots < \pi^*(i_K) \: .
$$
The algorithm computes the prices in the reverse order, starting from ad $i_K$.
First, it computes $p_{i_K}$ using the algorithm from
Section~\ref{section:pricing-algorithm}. Then, the bid of ad $i_K$ is lowered to
$p_{i_K}$. Note that the optimal assignment $\pi^*$ is still optimal even with
the lower bid. Second, the algorithm computes the price $p_{i_{K-1}}$ using
again the algorithm from Section~\ref{section:pricing-algorithm}. Similarly as
before, the bid of ad $i_{K-1}$ is lowered to $p_{i_{K-1}}$. The bid of ad $i_K$
is kept at $p_{i_K}$. Third, the algorithm computes the price $p_{i_{K-2}}$
using again the algorithm from Section~\ref{section:pricing-algorithm}. In
general, in round $r=1,2,\dots,K$, the algorithm computes the price
$p_{i_{K-r+1}}$ of ad $i_{K-r+1}$ using algorithm from
Section~\ref{section:pricing-algorithm} and it sets the bids of ads $i_K,
i_{K-1}, \dots, i_{K-r+1}$ to the prices computed previously. The lowered bids
are used in the successive rounds for advertiser $f$. The lowered bids have the
property that $\pi^*$ is the optimal assignment. Once prices for all ads from
advertiser $f$ are computed, all bids are restored to their original values
and the algorithm moves on to a different advertiser.

If each ad is from a different advertiser, the prices are identical to the
prices computed by the algorithm from Section~\ref{section:pricing-algorithm}.
The time complexity of the pricing algorithm is $O((N - M + 1)N_1 N_2)$ where
$M$ is the number of distinct advertisers with at least one ad in the optimal
assignment. In the worst case, when all ads are from the same advertiser,
time complexity is $O(NN_1 N_2)$.

\bibliographystyle{plainnat}
\bibliography{biblio}

\end{document}